\documentclass[sigplan,twocolumn]{acmart}
\renewcommand\footnotetextcopyrightpermission[1]{}
\usepackage{amsmath,amssymb,bbm,mathtools}
\usepackage{graphicx,subcaption,booktabs,multirow}
\usepackage{xspace,enumitem,microtype}
\usepackage{tikz}
\usepackage{hyperref}
\usepackage{microtype}
\usepackage{graphicx}
\usepackage{subcaption}
\usepackage{booktabs} 

\usepackage{enumitem}
\usepackage{microtype}
\usepackage{xspace}
\usepackage{multirow}
\usepackage{seqsplit}
\usepackage{xcolor}
\usepackage{hyperref}
\hypersetup{
    colorlinks=true,
    urlcolor=blue,
    linkcolor=blue,
    citecolor=blue
}
\newcommand{\sys}{LMDeploy\xspace}
\newcommand{\eng}{TurboMind\xspace}
\usepackage{xcolor}
\definecolor{lightgreen}{RGB}{144,238,144}
\definecolor{darkcyan}{HTML}{008B8B}

\usepackage{hyperref}
\usepackage{xurl}
\newcommand{\jyh}[1]{{\textcolor{black}{#1}}}

\usepackage{algorithm}
\usepackage{algorithmic}

\begin{document}

\title{\sys Accelerates Mixed-Precision\\ LLM Inference with \eng}

\author{Li Zhang}
\authornote{Both authors contributed equally to this research.}
\email{zhangli@pjlab.org.cn}
\affiliation{\institution{Shanghai AI Laboratory}\city{Shanghai}\country{China}}

\author{Youhe Jiang}
\authornotemark[1]
\authornote{This work is done during an internship at Shanghai AI Laboratory.}
\email{youhejiang@gmail.com}
\affiliation{\institution{Shanghai AI Laboratory}\city{Shanghai}\country{China}}

\author{Guoliang He}
\email{heguoliang@pjlab.org.cn}
\affiliation{\institution{Shanghai AI Laboratory}\city{Shanghai}\country{China}}

\author{Xin Chen}
\email{chenxin@pjlab.org.cn}
\affiliation{\institution{Shanghai AI Laboratory}\city{Shanghai}\country{China}}

\author{Han Lv}
\email{lvhan@pjlab.org.cn}
\affiliation{\institution{Shanghai AI Laboratory}\city{Shanghai}\country{China}}

\author{Qian Yao}
\email{yaoqian@pjlab.org.cn}
\affiliation{\institution{Shanghai AI Laboratory}\city{Shanghai}\country{China}}

\author{Ningsheng Ma}
\email{maningsheng@pjlab.org.cn}
\affiliation{\institution{Shanghai AI Laboratory}\city{Shanghai}\country{China}}

\author{Fangcheng Fu}
\email{ccchengff@sjtu.edu.cn}
\affiliation{\institution{Shanghai Jiao Tong University}\city{Shanghai}\country{China}}

\author{Kai Chen}
\email{chenkai@pjlab.org.cn}
\affiliation{\institution{Shanghai AI Laboratory}\city{Shanghai}\country{China}}

\begin{abstract}
Mixed-precision inference techniques reduce the memory and computational demands of Large Language Models (LLMs) by applying hybrid precision formats to model weights, activations, and KV caches. However, existing systems struggle to (\textbf{\underline{i}}) automatically generalize across diverse hardware architectures and precision formats, often requiring fragmented, hand-tuned kernels, and (\textbf{\underline{ii}}) fully exploit available memory and compute resources, often causing performance bottlenecks. To address these problems, we propose \eng, a \textbf{generalizable} and \textbf{efficient} mixed-precision LLM inference engine of \sys.
\eng is built around two hardware-aware mixed-precision pipelines: A General Matrix Multiply (GEMM) pipeline that optimizes matrix operations through offline weight packing and online acceleration, and an attention pipeline that enables efficient attention computation with different Query, Key, and Value precision combinations. These pipelines are enabled by four key techniques: (\textbf{\underline{i}}) Hardware-aware weight packing and (\textbf{\underline{ii}}) adaptive head alignment for \textbf{generalizability}, and (\textbf{\underline{iii}}) instruction-level parallelism and (\textbf{\underline{iv}}) a KV memory loading pipeline for \textbf{efficiency}.
We conduct comprehensive evaluations of \sys powered by \eng across sixteen popular LLMs and four representative GPU architectures. Results demonstrate that \sys achieves up to 61\% lower serving latency (30\% on average) and up to 156\% higher throughput (58\% on average) in mixed-precision workloads compared to existing mixed-precision frameworks, establishing consistent performance improvements across all tested configurations and hardware types. This work is open-sourced and publicly available at \url{https://github.com/InternLM/lmdeploy}.
\end{abstract}


\maketitle
\pagestyle{plain}

\section{Introduction}
\label{sec:intro}
Large language models (LLMs) such as DeepSeek-R1~\citep{guo2025deepseek}, OpenAI o3~\citep{gpt4o}, Claude~\citep{claude3}, Gemini~\citep{reid2024gemini} and Llama-3~\citep{dubey2024llama} have demonstrated outstanding performance across a wide range of applications (e.g., chatbots, healthcare and education)~\citep{jeon2023large,peng2023study,copilot}.
However, LLM inference and serving are costly~\cite{jiang2024hexgen,miao2024spotserve}, as the large number of model parameters and the generative nature of inference demand substantial memory and compute resources for efficient execution.
Recent efforts~\cite{lin2024awq,frantar2022gptq,lin2024qserve,xiao2023smoothquant,liu2024kivi} have explored quantization techniques for efficient LLM inference and serving. Typically, quantization reduces the precision of model weights, activations, and Key-Value (KV) caches from high-precision formats (e.g., FP16, FP32) to lower-precision representations (e.g., INT4, FP8), and carries out the inference process in a mixed-precision manner, significantly reducing memory footprint, memory I/O, and computational overhead~\cite{frantar2025marlin,hooper2024kvquant,dettmers2023qlora}. 
These techniques facilitate more efficient deployment of LLMs while maintaining acceptable model accuracy. 

Efficient mixed-precision inference requires extensive \seqsplit{hardware}-specific optimizations for different precision formats, in order to adapt to the multi-layer memory hierarchy and maximize the tensor core utilization of modern hardware.
Although recent works have focused on optimizing mixed-precision inference efficiency~\cite{frantar2025marlin,lin2024qserve,du2025bitdecoding,tensorrt-llm,pytorch}, we identify two fundamental optimization pillars---in which existing frameworks usually fall short---that are essential for delivering high-quality mixed-precision inference.

\vspace{0.5em}
\noindent \textbf{\underline{Pillar 1:} Configuration-agnostic automatic optimization.} To provide flexibility in balancing deployment efficiency and model accuracy based on user requirements and hardware constraints, mixed-precision inference systems necessitate consistent high-efficiency execution across diverse hardware architectures and precision formats~\cite{zhao2024atom,xiao2023smoothquant,frantar2022gptq,lin2024awq}. 
However, existing frameworks rely heavily on hardware- or format-specific optimization methods, lacking a unified abstraction for automatic adaptation. This rigidity demands significant engineering effort to generalize across configurations and often results in suboptimal performance. For instance, MARLIN~\cite{frantar2025marlin} is optimized specifically for the A100 architecture, while QServe~\cite{lin2024qserve} is tailored to the W4A8KV4\footnote{We use ``W$x$A$y$KV$z$'' to denote the mixed-precision format of $x$-bit weights, $y$-bit activations, and $z$-bit KV caches.} precision format. Both require substantial re-engineering to generalize to other hardware or precision configurations.

\vspace{0.5em}
\noindent \textbf{\underline{Pillar 2:} Efficient hardware utilization.}
To achieve efficient LLM deployment, mixed-precision inference systems necessitate thorough exploitation of modern GPU hardware resources, including the memory hierarchy (e.g., global, shared, and register memory), compute units (e.g.,  Arithmetic-Logic execution Units (ALUs) and tensor cores), and strategic overlapping of memory access with computation~\cite{li2024llm,jin2024comprehensive,daoflashattention2}.
However, existing frameworks typically fail to fully exploit these resources, resulting in underutilized hardware capabilities and suboptimal inference performance. For instance, TensorRT-LLM~\cite{tensorrt-llm} inadequately leverages the memory hierarchy, incurring substantial runtime overhead during dequantization~\cite{lin2024qserve}.

To achieve these two optimization pillars, we implement \eng, a \textbf{generalizable} and \textbf{efficient} mixed-precision LLM inference engine of LMDeploy. Our key contributions are summarized as follows:

\vspace{0.5em}
\noindent\textbf{\underline{Contribution 1:}} \textbf{Bottleneck analysis.} We systematically identify and analyze the key memory and compute optimization challenges in typical mixed-precision workflows. These encompass (\textbf{\underline{i}}) memory access inefficiencies including memory coalescing failures, bank conflicts, and misalignment issues across different hardware memory hierarchies (global memory, shared memory, register memory), as well as (\textbf{\underline{ii}}) computational bottlenecks including dequantization overhead, suboptimal tensor core utilization, and inefficient attention computation patterns.

\vspace{0.5em}
\noindent\textbf{\underline{Contribution 2:}} \textbf{\eng techniques.} We propose key techniques that address the identified bottlenecks while achieving both optimization pillars. For generalizability (Pillar 1), we introduce \textit{hardware-aware weight packing} and \textit{adaptive head alignment}, enabling automatic optimization across diverse hardware architectures and precision formats. For efficiency (Pillar 2), we employ \textit{instruction-level parallelism} to minimize dequantization overhead and a \textit{KV memory loading pipeline} to enhance attention computation throughput. These techniques are integrated into \eng for high-performance mixed-precision inference.

\vspace{0.5em}
\noindent\textbf{\underline{Contribution 3:}}  \textbf{Comprehensive evaluation.} We conduct comprehensive evaluations of \sys powered by \eng across 16 popular LLMs spanning both dense and Mixture-of-Experts (MoE) architectures as well as 4 representative GPU architectures (RTX 4090, L40S, A100, H100), and comparing against existing mixed-precision systems including vLLM enhanced by MARLIN~\cite{vllm_quantized_kvcache,kwon2023efficient,frantar2025marlin}, TensorRT-LLM~\cite{tensorrt-llm}, and OmniServe with integrated QServe optimization~\cite{lin2024qserve}. Results demonstrate that our system consistently achieves significant performance improvements across all tested configurations, delivering up to 61\% lower serving latency (30\% on average) and up to 156\% higher throughput (58\% on average) compared to existing frameworks.

The remainder of this paper is organized as follows. 
\S\ref{sec:background} provides background and discusses related work.
\S\ref{sec:quantizationpipeline} characterizes the key memory and compute bottlenecks in mixed-precision inference and introduces the overall design of \eng. 
\S\ref{sec:turbomind} presents the core system techniques that address these bottlenecks.
\S\ref{sec:eval} evaluates LMDeploy across representative models, workloads, precision formats, and GPU architectures.
\S\ref{sec:futurework} discusses future work.
\S\ref{sec:conclusion} concludes the paper.

\section{Background and Related Works}
\label{sec:background}

\noindent\textbf{Memory hierarchy of modern GPUs.} Modern GPU architectures feature complex memory hierarchies with distinct performance characteristics that significantly impact LLM inference efficiency. Contemporary GPUs such as the A100 provide multiple memory types including high-bandwidth memory (HBM), L2 cache, shared memory, and register files, each with different latency, bandwidth, and capacity trade-offs~\cite{nvidia_a100}. Recent research has explored this hierarchy for LLM workloads~\cite{dao2022flashattention,daoflashattention2,aminabadi2022deepspeed,NVIDIA2019FasterTransformer,fang2021turbotransformers,nvidia_cutlass,nvidia_cudnn,chen2018tvm,zheng2020ansor}. FlashAttention~\cite{dao2022flashattention} and FlashAttention-2~\cite{daoflashattention2} demonstrate substantial improvements through careful orchestration of memory accesses and intermediate result placement; DeepSpeed-Inference~\cite{aminabadi2022deepspeed} and FasterTransformer~\cite{NVIDIA2019FasterTransformer} leverage memory hierarchy awareness for kernel design and data placement strategies; TVM~\cite{chen2018tvm} and Ansor~\cite{zheng2020ansor} enable compiler-level memory hierarchy optimization through automated scheduling; and CUTLASS~\cite{nvidia_cutlass} and cuDNN~\cite{nvidia_cudnn} provide memory-optimized GEMM kernels for different hierarchy levels.

\vspace{0.5em}
\noindent\textbf{Quantization techniques.} Quantization techniques have emerged as essential strategies for reducing the computational and memory demands of LLM inference~\cite{lin2024awq,frantar2022gptq,xiao2023smoothquant,liu2024kivi,hezipcache,kang2024gear}.
Weight quantization approaches such as GPTQ~\cite{frantar2022gptq}, AWQ~\cite{lin2024awq}, and SmoothQuant~\cite{xiao2023smoothquant} demonstrate that 4-bit and 8-bit model weights can maintain acceptable accuracy while significantly reducing memory footprint.
Additionally, KV cache quantization methods focus on compressing the cache during inference. KIVI~\cite{liu2024kivi} applies asymmetric 2-bit quantization to KV cache, achieving significant memory reduction while preserving inference quality; 
ZipCache~\cite{hezipcache} performs accurate and efficient KV cache quantization with salient token identification;
and GEAR~\cite{kang2024gear} integrates uniform quantization, low-rank matrix approximation, and sparse matrix handling for near-lossless generative inference.

\vspace{0.5em}
\noindent\textbf{Performance optimization for mixed-precision inference.} Recent frameworks aim to improve mixed-precision performance and maximize inference efficiency~\cite{tensorrt-llm,pytorch,frantar2025marlin,lin2024qserve,lin2024qserve}. MARLIN~\cite{frantar2025marlin} provides a highly optimized GPU kernel design that enables near-optimal inference performance for quantized LLMs; TensorRT-LLM~\cite{tensorrt-llm} offers comprehensive mixed-precision optimization through graph-level transformations; QServe~\cite{lin2024qserve} introduces novel quantization-aware serving algorithms that co-optimize model compression with system throughput.
Recent research also focuses on accelerating low-bit KV cache inference~\cite{vllm_quantized_kvcache,du2025bitdecoding}. vLLM~\cite{kwon2023efficient} implements FP8 KV cache support and specialized attention kernels for reduced precision operations~\cite{vllm_quantized_kvcache}; and BitDecoding~\cite{du2025bitdecoding} targets low-bit KV cache compression to unlock tensor core utilization for long-context LLM decoding scenarios. 
However, there are two major challenges within current frameworks: (\textbf{\underline{i}}) They fail to fully exploit the memory hierarchy and tensor core resources of modern hardware. For instance, TensorRT-LLM suffers from significant runtime dequantization overhead with INT4 quantization~\cite{lin2024qserve}; (\textbf{\underline{ii}}) They lack holistic support and optimization for different mixed-precision inference formats. For instance, MARLIN supports GEMM kernel optimization only~\cite{frantar2025marlin}, and QServe is hard-wired to the W4A8KV4 precision format~\cite{lin2024qserve}.

\vspace{0.5em}
\noindent\textbf{LLM inference and serving.} The rapid deployment of LLMs in production environments has driven significant research into efficient inference and serving systems~\cite{yu2022orca,li2023alpaserve,zheng2024sglang,kwon2023efficient,zhong2024distserve,patel2024splitwise,jiang2024hexgen,jiang2025thunderserve,agrawal2024taming,tensorrt-llm}. 
Among them, TensorRT-LLM emphasizes highly optimized CUDA kernels and graph-level optimizations~\cite{tensorrt-llm};
AlpaServe~\cite{li2023alpaserve} adpots model parallelism to optimize LLM serving performance;
vLLM~\cite{kwon2023efficient} introducing PagedAttention for dynamic memory management and continuous batching;
SGLang~\cite{zheng2024sglang} advances LLM serving by co-designing the frontend language with the serving backend to enable more efficient execution of complex LLM programs;
Splitwise~\cite{patel2024splitwise} and DistServe~\cite{zhong2024distserve} explore distributed serving strategies that disaggregate prefill and decoding phases across different resource pools to optimize resource utilization; 
SarathiServe~\cite{agrawal2024taming} introduces a chunked prefill approach and piggybacks decoding requests to improve hardware utilization; and HexGen~\cite{jiang2024hexgen} and ThunderServe~\cite{jiang2025thunderserve} propose to use heterogeneous resources for cost-efficient LLM serving.

\section{Challenges and \eng Design}
\label{sec:quantizationpipeline}

This section first introduces a typical mixed-precision inference workflow (\S\ref{subsec:typical}), then discusses the key challenges in mixed-precision workflows (\S\ref{sec:memory challenges}), and finally presents the design of \eng (\S\ref{sec:pipelines}).

\begin{figure}[t!]
    \centering
    \includegraphics[width=\linewidth]{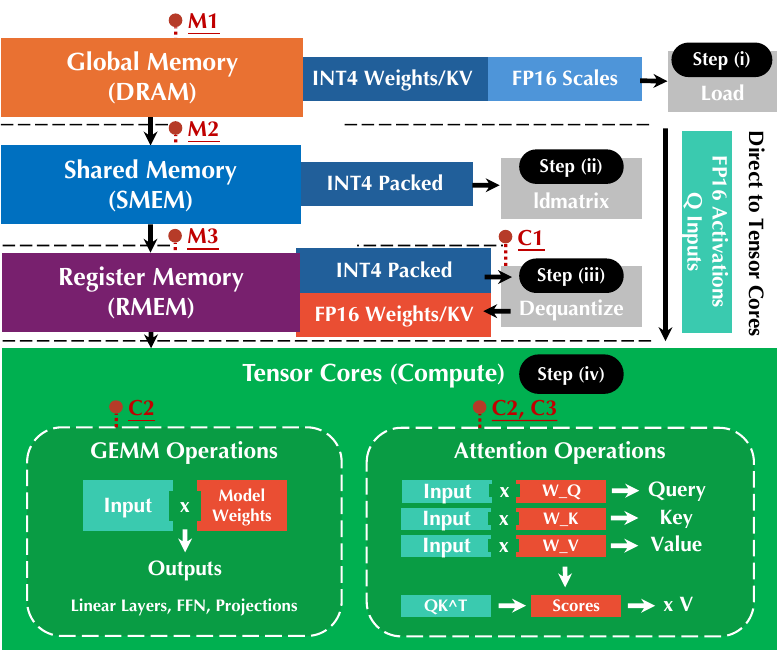}
    \caption{Illustration of the memory hierarchy and each step of the mixed-precision inference workflow.}
    \label{fig:workflow}
\end{figure}

\subsection{Typical Mixed-Precision Inference Workflow}
\label{subsec:typical}
\autoref{fig:workflow} illustrates the memory hierarchy and five essential steps in a typical mixed-precision inference workflow with INT4 model weights and KV cache:
(\textbf{\underline{i}}) Loading INT4 quantized weights and FP16 scales from global memory into shared memory; (\textbf{\underline{ii}}) transferring the packed INT4 values to registers using hardware instructions (e.g., \texttt{ldmatrix}); (\textbf{\underline{iii}}) dequantizing the INT4 values to FP16 format through bit manipulation operations, and applying the quantization scales; (\textbf{\underline{iv}}) feeding the dequantized FP16 weights into tensor cores alongside the FP16 activations to perform standard FP16 matrix multiplication.
However, each operation (e.g., memory loading and computation) in the workflow must be carefully managed to achieve efficient implementation.

\subsection{Memory and Compute Optimization Challenges}
\label{sec:memory challenges}
\label{sec:compute challenges}

Both memory loading and computation can become bottlenecks in mixed-precision LLM inference~\cite{agrawal2024taming,patel2024splitwise,zhong2024distserve,frantar2025marlin}. We list the key optimization challenges of the mixed-precision inference workflow, categorized into memory (\underline{M1}-\underline{M3}) and computation (\underline{C1}-\underline{C3}), as follows.

\vspace{0.5em}
\noindent\textbf{\underline{M1:} Global memory coalescing.}
Modern GPUs achieve peak memory bandwidth when the memory addresses accessed by every thread within a warp are within the same aligned segment of global memory (e.g., 3-byte on Hopper/Ampere). This alignment enables the warp to access contiguous memory regions through one efficient global memory transaction~\cite{alur2017gpudrano,fauzia2015characterizing,kim2017evaluation}.
In mixed-precision inference, however, packing weights into low-bit formats results in misalignment between each warp's memory accesses and the GPU's standard 32-/64-/128-byte memory segments~\cite{cuda_prog_guide_128b_2024,nvidia_tensorrt_guide_2024,pytorch_gptq_optimization}. This misalignment necessitates multiple global memory transactions per warp for operations rather than one efficient transaction, thereby significantly reducing effective memory bandwidth~\cite{kim2024quick,frantar2022gptq}. An illustration of this process is shown in~\autoref{appendix:challenges}.

\vspace{0.5em}
\noindent\textbf{\underline{M2:} Shared memory bank conflicts.}
Shared memory bank conflicts occur when multiple threads within a warp simultaneously access different addresses that map to the same memory bank, forcing these accesses to be serialized rather than executed in parallel, thereby reducing memory throughput by up to an order of magnitude~\cite{biswas2020memory,horga2022symbolic,gao2014improving,lin2024qserve}. In mixed-precision inference, low-bit values are typically packed into larger-bit words (e.g., eight INT4 values per 32‑bit word~\cite{nvidia_tensorrt_quant}) to achieve optimal memory bandwidth and space efficiency~\cite{frantar2025marlin,rakka2022mixed,pytorch,tensorrt-llm}. 
However, packing low‑bit weights into larger-bit words makes each column load jump a full packed row (e.g., 128-byte for 32 threads read 32‑bit words)~\cite{nvidia_tensorrt_guide_2024,cuda_prog_guide_128b_2024}. That stride maps every thread in the warp to the same shared memory bank, so the bank must serve the requests one‑by‑one, turning a parallel-access load into a serialized‑access stall, severely degrading memory throughput. An illustration of this process is shown in~\autoref{appendix:challenges}.

\begin{figure}
    \centering
    \includegraphics[width=\linewidth]{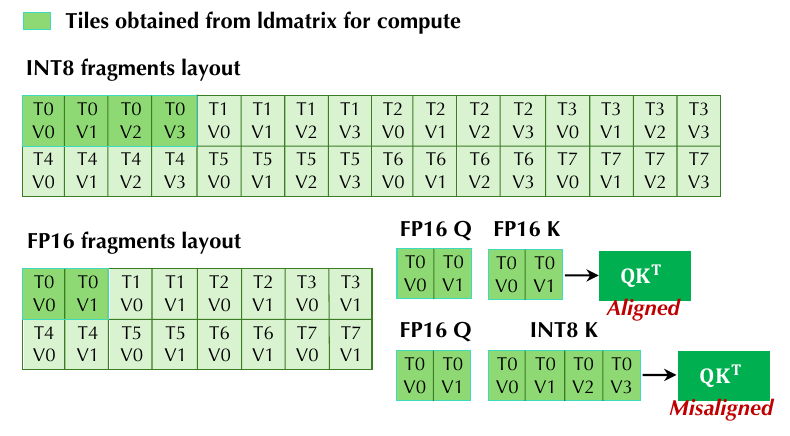}
    \caption{Illustration of register memory misalignment.}
    \label{fig:registermemory}
\end{figure}

\vspace{0.5em}
\noindent\textbf{\underline{M3:} Register memory misalignment in low-bit KV cache inference.} In mixed-precision inference, compressing the KV cache to low-bit precision (e.g., INT4 or INT8) while leaving the query matrix $Q$ in FP16 creates a byte-stride mismatch: warp-level matrix-load instructions such as \texttt{ldmatrix} fetch wider tiles per lane for $K$ than for $Q$~\cite{du2025bitdecoding,nvidia_ldmatrix_isa,tan2024alignedkv}. The tensor core therefore multiplies misaligned fragments and produces incorrect attention scores in $S = QK^{\mathsf T}$, as shown in~\autoref{fig:registermemory}. 
Current practices mainly disable the warp-level matrix-load instruction to eliminate the layout mismatch. However, this approach requires tensor-core tile reconstruction through additional per-lane address arithmetic and shuffle operations, thereby incurring additional computational overhead and negating most throughput benefits of low-bit KV-cache inference~\cite{kim2024quick,du2025bitdecoding,luitjens2013vector}.

\vspace{0.5em}
\noindent\textbf{\underline{C1:} Dequantization overhead.} Modern GPUs lack native hardware support for mixed-precision arithmetic between low-bit and FP16 operands, necessitating dequantization before MMA and attention computation~\cite{kim2024quick}. However, naïve type casts from low-bit to FP16 are slow and can become a computational bottleneck~\cite{kim2022says,frantar2025marlin}, making efficient dequantization essential for mixed-precision inference.

\begin{figure}[t!]
    \centering
    \includegraphics[width=\linewidth]{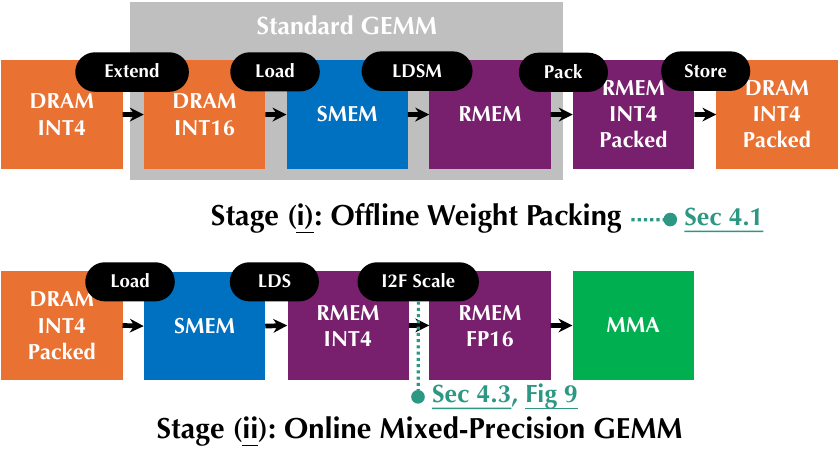}
    \caption{Illustration of GEMM pipeline design. LDSM and LDS are shared memory load instructions for matrix and non-matrix data on Ampere, which can be changed to other specific instructions on other GPU generations. I2F denotes Integer-to-Float conversion for dequantizing low-bit values to FP16.}
    \label{fig:gemm pipeline}
\end{figure}

\vspace{0.5em}
\noindent\textbf{\underline{C2:} MMA data misalignment with quantization.} 
Modern tensor cores provide INT8/INT4 MMA instructions, but they only run at full throughput when the inputs are pre-packed into fixed tiles (e.g., 16$\times$8$\times$32/64 for Ampere, \seqsplit{16$\times$8$\times$64/128} for Hopper)~\cite{nvidia_ampere_tuning_guide,luo2024benchmarking}.
Standard quantization layouts seldom meet these hardware requirements, creating a fundamental mismatch between them. Consequently, kernels must either inject padding that wastes compute resources, perform costly in-register shuffles at runtime, or fall back to less efficient scalar operations—all of which erode tensor core throughput \cite{lin2024awq,frantar2025marlin}.

\vspace{0.5em}
\noindent\textbf{\underline{C3:} Attention computation bubbles.} During the decoding phase, each attention head must first load the new key/value rows from the KV cache before the tensor core can start the dot product, causing stall windows since memory traffic and arithmetic execution are serialized rather than overlapped~\cite{patel2024splitwise,zhong2024distserve}. Quantized KV caches exacerbate this issue: low-bit keys must be fetched and dequantized to FP16 in registers before being consumed by MMA, further extending the stall windows~\cite{du2025bitdecoding,kim2024quick}. Thus, the theoretical memory-bandwidth savings from quantization are often negated by the increased bubbles caused by dequantization, yielding negative performance gains in attention-heavy workloads.

\subsection{\eng Design}
\label{sec:pipelines}

To address the above challenges, we propose two pipeline designs for GEMM and attention operations in \eng, optimizing memory loading and computation coordination across the memory hierarchy.

\begin{figure}[t!]
    \centering
    \includegraphics[width=\linewidth]{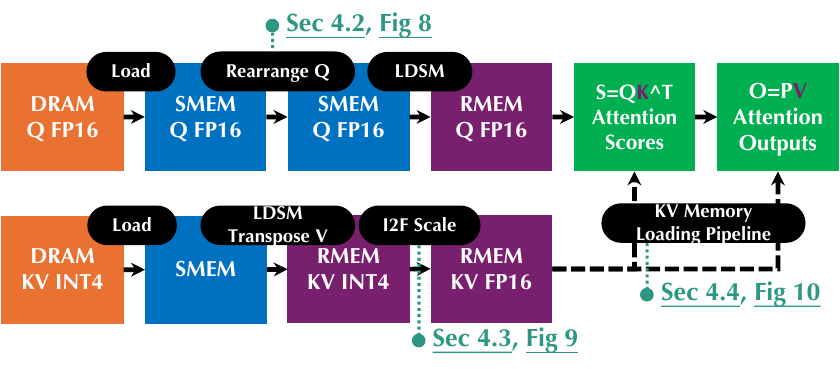}
    \caption{Illustration of attention pipeline design. The transpose V operation converts V to a column-major tile layout for tensor core compatibility, and the final output O is rearranged back into row-major linear memory before the global write.}
    \label{fig:attention pipeline}
\end{figure}

\vspace{0.5em}
\noindent\textbf{GEMM pipeline design.} \autoref{fig:gemm pipeline} illustrates the GEMM pipeline design of \eng, which is divided into two stages: (\textbf{\underline{i}}) Offline weight packing and (\textbf{\underline{ii}}) online mixed-precision GEMM.
In the offline stage, we employ \textit{hardware-aware weight packing} (\S\ref{sec:hardware-aware}) to convert weight matrices into layouts optimized for efficient memory access and computation, eliminating global memory coalescing issues, bank conflicts, and MMA data misalignment (\underline{M1}-\underline{2}, \underline{C2}).
In the online stage, pre-processed low-bit fragments are loaded through the standard memory hierarchy and dequantized to FP16 using \textit{instruction-level parallelism} (\S\ref{sec:ilp}) to minimize dequantization overhead (\underline{C1}). Through offline weight packing, the online stage requires no additional runtime transformations for efficient mixed-precision inference.

\begin{figure}[t!]
    \centering
    \includegraphics[width=\linewidth]{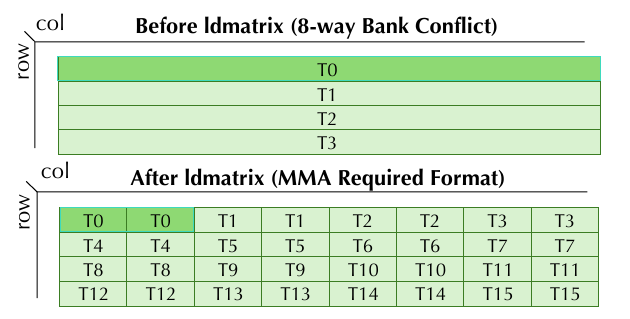}
    \caption{Illustration of matrix-load instructions' redistribution in step (\textbf{\underline{ii}}). Before \texttt{ldmatrix}, each thread is responsible for loading one matrix row (16-byte), resulting in 8-way bank conflict. This problem is resolved after \texttt{ldmatrix}.}
    \label{fig:step2}
\end{figure}

\vspace{0.5em}
\noindent\textbf{Attention pipeline design.} \autoref{fig:attention pipeline} illustrates the attention pipeline design of \eng, which processes Q and KV through separate yet coordinated branches.
In the Q branch, we employ \textit{adaptive head alignment} (\S\ref{sec:adaptive-head}) to rearrange Q in shared memory, resolving register memory misalignment in low-bit KV cache inference (\underline{M3}).
In the KV branch, low-bit KV caches are loaded from global memory, moved through the standard memory hierarchy, and dequantized to FP16 using \textit{instruction-level parallelism} (\S\ref{sec:ilp}) to minimize dequantization overhead (\underline{C1}).
Furthermore, the \textit{KV Memory Loading Pipeline} (\S\ref{sec:klo}) overlaps KV memory loading with dequantization and attention computation, minimizing attention computation bubbles (\underline{C3}).

\section{\eng Techniques}
\label{sec:turbomind}

This section presents four core techniques that instantiate the GEMM and attention pipeline designs of \eng, including hardware-aware weight packing (\S\ref{sec:hardware-aware}) and adaptive head alignment (\S\ref{sec:adaptive-head}), instruction-level parallelism (\S\ref{sec:ilp}), and the KV memory loading pipeline (\S\ref{sec:klo}). Throughout this section, we highlight how \eng differs from existing mixed-precision inference frameworks.

\subsection{Hardware-aware Weight Packing}
\label{sec:hardware-aware}

Current weight packing approaches typically employ \textit{static} weight layout designs that are optimized for specific hardware configurations and tensor core instruction sets~\cite{tensorrt-llm,lin2024qserve}. For instance, MARLIN requires manual specifications of optimal configurations including warp layouts and tile sizes, necessitating extensive tuning for different GPU architectures~\cite{frantar2025marlin}. To address this limitation, we propose a \textit{hardware-aware weight packing} technique that automatically adapts to different GPU architectures and MMA instruction requirements while maintaining strong performance across diverse hardware configurations. The hardware-aware weight packing process is performed entirely offline.

\begin{figure}[t!]
    \centering
    \includegraphics[width=\linewidth]{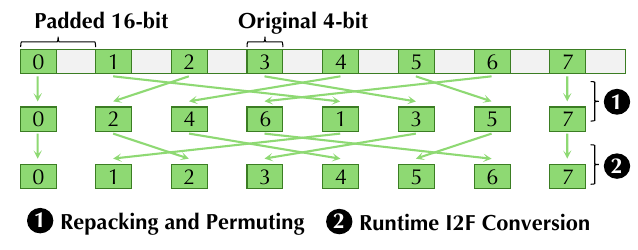}
    \caption{Illustration of repacking and permuting operations in step (\textbf{\underline{iii}}), and the runtime I2F conversion. The values \{0-7\} represent the indices of eight elements within a single thread fragment. This procedure guarantees that, after I2F conversion, the data already match the lane layout required by the MMA instruction.}
    \label{fig:step3}
\end{figure}

\begin{figure}[t!]
    \centering
    \includegraphics[width=\linewidth]{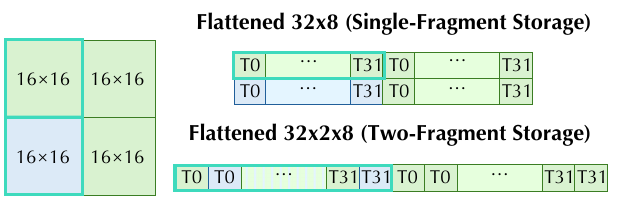}
    \caption{Illustration of fragment storage in step (\textbf{\underline{iv}}). Single‑ and two‑fragment storage refer to how many packed fragments are written in one store operation. We typically use two-fragment storage for LDS efficiency.}
    \label{fig:storing}
\end{figure}

\vspace{0.5em}
\noindent\textbf{{Hardware-aware weight packing steps.}} 
The key insight of our approach is to leverage existing higher precision memory-to-register data pipelines rather than design manual specifications for low-precision formats. 
During the offline weight packing process, we allow data pipelines to guide the layout transformation, producing packed weights that align perfectly with the hardware's memory hierarchy and tensor core requirements for efficient online inference.
Our approach consists of four essential steps:
\textbf{(\underline{i})} \textit{Bit extension.} The low-bit weights are temporarily widened to 16-bit format to ensure compatibility with standard (non-mixed-precision) GEMM pipelines. \textbf{(\underline{ii})} \textit{Fragment loading.} Each warp issues an asynchronous copy (e.g., \texttt{cp.async}) to move one cache-line-sized slice of the weight matrix (e.g., 128-byte on Ampere~\cite{cuda_prog_guide_128b_2024}) from global to shared memory, then invokes the matrix-load instruction (e.g., LDSM on Ampere) to load the slice into registers. In this step, the instruction's internal crossbar automatically redistributes words across lanes~\cite{gebhart2012unifying}, as shown in~\autoref{fig:step2}.
\textbf{(\underline{iii})} \textit{Bit compression.} 
Inside registers, the padded
16-bit words are repacked into their original low-bit format,
preserving the lane-level MMA layout \jyh{established in the previous step}, while permuting the sub-word values into the exact order expected by the MMA instruction, as shown in~\autoref{fig:step3}.
\textbf{(\underline{iv})} \textit{Fragment storing.} Each warp writes its packed fragments (typically two at a time for LDS efficiency) back to global memory with a single, fully coalesced cache-line store. As shown in~\autoref{fig:storing}, two 16$\times$16 fragments are stored in a flattened 32$\times$2$\times$8 format. This operation transforms the fragments into a contiguous layout for direct loading, thereby eliminating the additional swizzling operations required during runtime inference (detailed in ~\autoref{appendix:swizzle}).

Collectively, steps (\textbf{\underline{i}})-(\textbf{\underline{iv}}) pack the model weights into a hardware-optimized layout during the offline process (i.e., hardware-aware weight packing). During the online process (i.e., runtime inference), every warp can reload the weights directly and efficiently with the same two-instruction sequence from step (\textbf{\underline{ii}}): An asynchronous copy followed by the matrix-load instruction (e.g., \texttt{cp.async} + LDS on Ampere), without any additional alignment or swizzling overhead. Additionally, this design reduces contention for the ALU units that handle I2F conversions, as fewer arithmetic instructions (e.g., add, multiply) compete for the same computational resources, improving overall I2F efficiency.

\vspace{0.5em}
\noindent\textbf{Key advantages.} This technology optimizes weight packing across GPU architectures through a \textbf{unified procedure}, providing performance guarantees: Coalesced global memory transactions (\underline{M1}), bank-conflict-free shared memory access (\underline{M2}), and MMA data alignment (\underline{C2}). 
Consequently, our implementation achieves excellent performance, matching hand-tuned kernels without per-architecture retuning.
\jyh{We evaluate our GEMM kernel performance in~\S\ref{sec:kernelbench} and examine our GEMM pipeline design across different GPU generations in ~\S\ref{sec:e2e} to demonstrate the effectiveness and general applicability of our hardware-aware weight packing approach.}

\subsection{Adaptive Head Alignment}
\label{sec:adaptive-head}
As discussed in~\S\ref{sec:memory challenges}, mixing FP16 Q with low-bit KV misaligns warp fragments and corrupts $S = QK^{\mathsf T}$. 
Existing frameworks~\cite{pytorch,tensorrt-llm,vllm_quantized_kvcache} such as PyTorch, TensorRT, and vLLM dequantize the low-bit KV cache back to FP16 before matrix-load instructions to avoid misalignment. However, this extra conversion increases memory traffic and leaves the tensor cores idle, which lowers overall utilization. In contrast to existing works, \jyh{\eng performs a \textit{lightweight} rearrangement of the FP16 Q tensor once per decoding step, aligning its warp‑level loads with the low‑precision K tiles}—thereby delivering satisfactory tensor core throughput without sacrificing utilization.

\begin{figure}[t!]
    \centering
    \includegraphics[width=\linewidth]{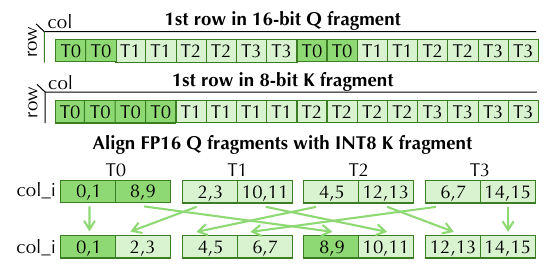}
    \caption{Illustration of aligning FP16 Q with INT8 K fragment. col\_i represents the column index.}
    \label{fig:match}
\end{figure}

\vspace{0.5em}
\noindent\textbf{Rearrange Q.} We rearrange Q to different KV precisions in three steps: (\textbf{\underline{i}}) Computing the appropriate number of K-slices of the Q matrix based on K matrix precision, where K-slices are chunks of the Q matrix along the K dimension. For example, 128-dimensional Q heads require 8, 16, and 32 K-slices for FP16, INT8, and INT4 operands respectively ($\text{OP\_K}$=16, 8, 4). This step adapts Q slicing to different K matrix precisions for fragment compatibility.
(\textbf{\underline{ii}})
Coordinating thread mapping to Q elements in shared memory access patterns. Each of the 32 threads within a warp computes unique row and column indices to target distinct Q matrix elements. This step ensures conflict-free shared memory bank access during concurrent Q loading.
(\textbf{\underline{iii}}) Employing the LDS instruction to rearrange Q matrix elements into register layouts compatible with tensor cores. The instruction loads Q values from shared memory, and it redistributes them across the 32 threads within a warp (similar to the fragment loading step in~\S\ref{sec:hardware-aware}).
This step enables efficient tensor core operations without fragment misalignment penalties. 
We demonstrate the illustration of aligning FP16 Q with INT8 K fragment in~\autoref{fig:match}.
We also present the detailed rearrangement pseudocode in Algorithm~\autoref{alg:transformQ} in~\autoref{appendix:rearrangement}.

\vspace{0.5em}
\noindent\textbf{Key advantages.} This technology offers two benefits: (\textbf{\underline{i}}) Seamless adaptation to arbitrary KV precisions through a \textbf{unified} rearrangement \textbf{procedure}, and (\textbf{\underline{ii}}) minimal overhead, as the rearrangement occurs only \textit{once} per attention head during shared memory-to-register loading. These properties resolve register memory misalignment (\underline{M3}). We evaluate this design in~\S\ref{sec:e2e} and~\S\ref{sec:kernelbench}.

\subsection{Instruction-level Parallelism}
\label{sec:ilp}
Simply reading low-bit weights into registers and performing explicit I2F casts is slow (\S\ref{sec:compute challenges}), creating a computation bottleneck in the mixed-precision inference workflow. Thus, we exploit \textit{instruction-level parallelism (ILP)} to minimize dequantization overhead.

\begin{figure}[t!]
    \centering
    \includegraphics[width=\linewidth]{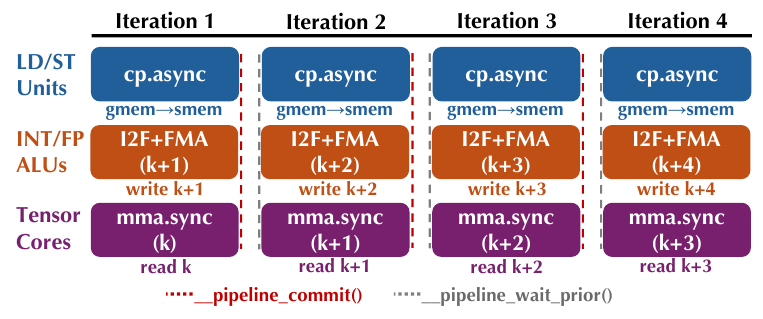}
    \caption{Overall process of parallel MMA-dequantization.}
    \label{fig:three-way}
\end{figure}

\vspace{0.5em}
\noindent\textbf{Parallel MMA-dequantization.} To minimize the dequantization overhead, we implement a software-pipelined mainloop that orchestrates three concurrent \jyh{stages} across different execution units: (\textbf{\underline{i}}) Tensor cores execute \texttt{mma.sync} operations on the current tile $k$, performing the matrix multiplication using previously dequantized fragments.
(\textbf{\underline{ii}}) INT/FP ALUs run the I2F conversion and Fused Multiply–Add (FMA) on tile $k+1$ while tensor cores are occupied.
(\textbf{\underline{iii}}) LD/ST (load/store execution) units asynchronously prefetch subsequent tiles\footnote{The number of prefetched tiles is determined by the configured memory pipeline depth, typically $\geq$ 3 on SM80+ GPU architectures~\cite{colfax_cutlass_tutorial_gemm_2024}.} from global memory using \texttt{cp.async}, preparing data for future iterations. This three-way overlap is enabled by strategic register allocation, which writes dequantized fragments to registers consumed by the next MMA iteration, eliminating read-after-write hazards. 
Additionally, CUDA's asynchronous pipeline primitives (\texttt{\seqsplit{pipeline\_commit}} and \texttt{pipeline\_wait\_prior}) manage overlapping \texttt{cp.async} operations, ensuring optimal load-compute synchronization throughout the pipeline.
We demonstrate the overall process in~\autoref{fig:three-way}.

\vspace{0.5em}
\noindent\textbf{Key advantages.} The instruction-level parallelism optimization can effectively eliminates the dequantization bottleneck (\underline{C1}). The parallel MMA-dequantization approach overlaps tensor core computation, I2F conversion and FMA, and \texttt{cp.async} operations. This technique enables low-bit matrix multiplication to approach the throughput of pure FP16$\times$FP16 kernels for large-batch inference, while outperforming them at smaller batch sizes, thereby enabling efficient mixed-precision inference across diverse workload scenarios. We perform the evaluation of our instruction-level parallelism in~\S\ref{sec:kernelbench}.

\subsection{KV Memory Loading Pipeline}
\label{sec:klo}
To reduce attention overhead during mixed-precision inference with low-bit KV cache, we implement a \textit{KV memory loading pipeline} that overlaps memory loading with attention computation, maximizing hardware bandwidth and compute utilization.

\begin{figure}[t!]
    \centering
    \includegraphics[width=\linewidth]{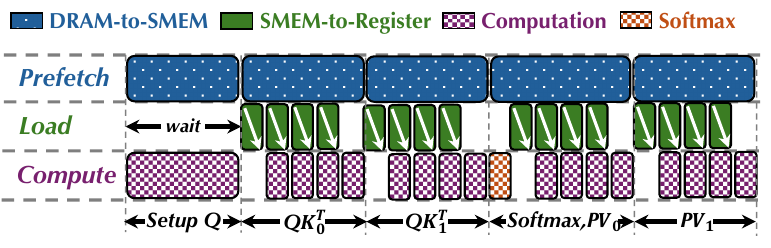}
    \caption{Illustration of the KV memory loading pipeline when the context spans two KV tiles ($K_0$, $V_0$ and $K_1$, $V_1$). The kernel executes the load–compute pipeline in 16-value micro-tiles (a macro-tile consists of 64 tokens)~\cite{colfax_cutlass_tutorial_gemm_2024}.}
    \label{fig:memoryloading}
\end{figure}

\vspace{0.5em}
\noindent\textbf{Triple-level parallelism.}
The KV memory loading pipeline orchestrates the concurrent loading of K- and V-tiles with $QK^{\mathsf T}$ and $PV$ computation\footnote{For low-bit KV inference, this loading step includes an additional I2F conversion to dequantize the cache to FP16.}.
Achieving this requires triple-level parallelism: (\textbf{\underline{i}}) Executing the current matrix multiply $QK^{\mathsf{T}}$/$PV$ on the 16-value slice of the K-/V-tile already in registers; (\textbf{\underline{ii}}) loading the next 16-value slice of that same tile from shared memory into registers (dequantizing on the fly for a low-bit KV cache); and (\textbf{\underline{iii}}) prefetching the next K-/V-tile from global to shared memory (this copy starts only when a memory pipeline stage is released).
\autoref{fig:memoryloading} demonstrates this coordination: While the tensor cores compute $QK_0^{\mathsf{T}}$ using the first 16-value slice of $K_0$, the pipeline simultaneously loads the second slice of $K_0$ into registers and prefetches the next tiles ($K_1$/$V_1$) from global memory.
These three overlapping stages keep the pipeline saturated, maximizing memory bandwidth utilization and effectively eliminating computation bubbles.

\vspace{0.5em}
\noindent\textbf{Key advantages.} This technique overlaps global-to-shared prefetching, shared-to-register loading, and tensor core computation to eliminate attention computation bubbles (\underline{C3}). It enables the attention kernel to sustain near-peak HBM bandwidth utilization across different KV cache precisions. We demonstrate the performance of our attention kernel in terms of latency and memory bandwidth utilization in~\S\ref{sec:kernelbench}.

\section{Evaluation}
\label{sec:eval}

In this section, we conduct a comprehensive comparison of \sys with state-of-the-art mixed-precision inference and serving frameworks. We seek to answer the following questions:
\begin{itemize}[leftmargin=*]
    \item (\textbf{\underline{Q1}}) \textit{How does LMDeploy compare with state-of-the-art mixed-precision inference frameworks in end-to-end serving performance across different models, hardware platforms, and workloads?}
    \vspace{0.25em}
    \item (\textbf{\underline{Q2}}) \textit{How effective are the proposed GEMM and attention pipeline designs at the kernel level?}
    \vspace{0.25em}
    \item (\textbf{\underline{Q3}}) \textit{How does LMDeploy perform under different KV cache precisions and multi-GPU serving configurations?}
\end{itemize}

\begin{figure*}[t!]
    \centering
    \includegraphics[width=\linewidth]{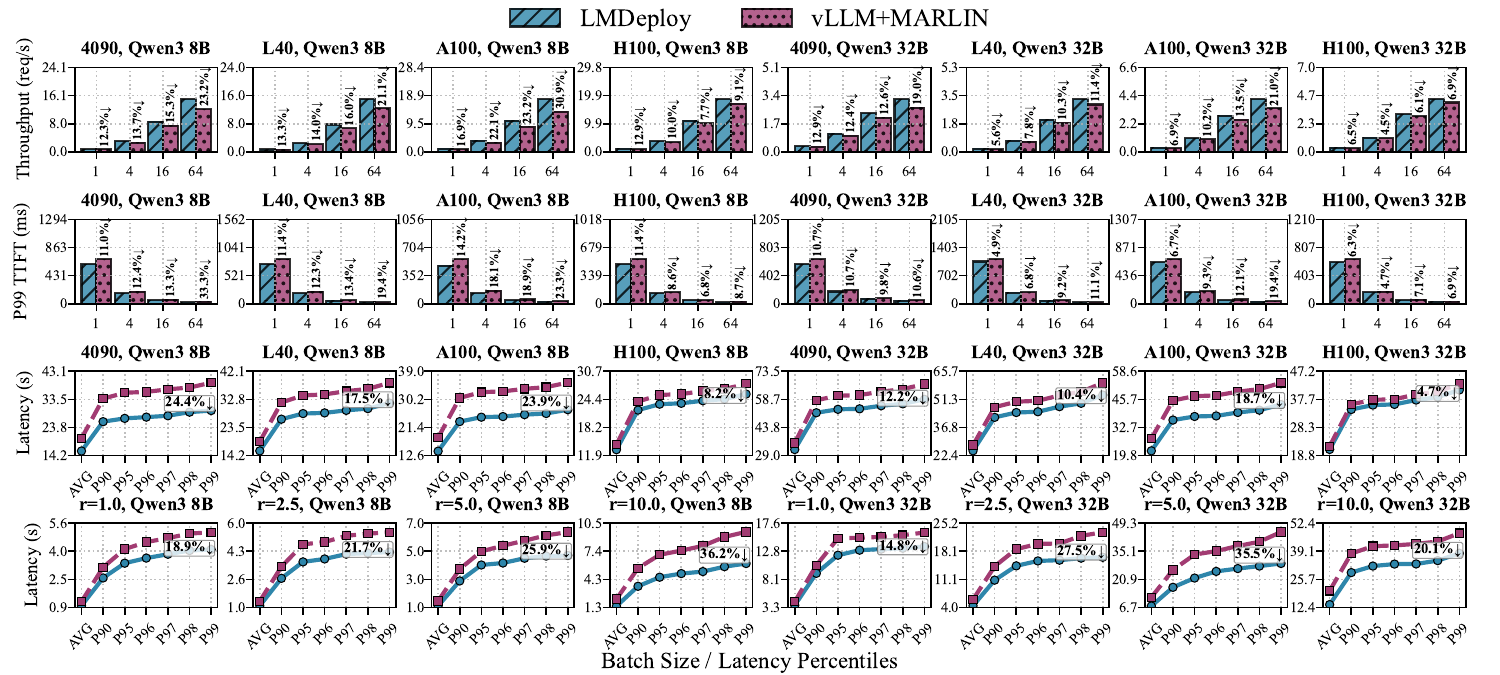}
    \caption{Compare \sys with vLLM+MARLIN. Rows show: (1-2) Throughput and TTFT latency across batch sizes, (3) latency for online serving at maximum batch size and request rate, and (4) latency under varying request rates on A100 GPU.}
    \label{fig:overall}
\end{figure*}

\subsection{Experiment Setup}
\textbf{Hardware environments.} Our experiments are conducted on four different GPU types: RTX 4090~\cite{nvidia_rtx4090}, L40S~\cite{nvidia_l40s}, A100 \cite{nvidia_a100}, and H100~\cite{nvidia_h100}. For ultra-large LLMs (e.g., Mixtral 8$\times$22B \cite{mistral_mixtral8x22b}, Qwen 235B~\cite{yang2025qwen3}), we utilize tensor parallelism~\cite{shoeybi2019megatron} to accommodate the large model size.

\vspace{0.5em}
\noindent\textbf{Baselines.} We compare \sys with other state-of-the-art mixed-precision inference frameworks:
\begin{itemize}[topsep=5pt, leftmargin=*]
\item \textbf{vLLM+MARLIN.} vLLM~\cite{kwon2023efficient} is a state-of-the-art serving framework integrated with MARLIN kernels~\cite{frantar2025marlin} for mixed-precision LLM inference. 
We utilize the latest version of vLLM (v0.9.1) for comparison.
\vspace{0.25em}
\item \textbf{TensorRT-LLM.} TensorRT-LLM~\cite{tensorrt-llm} is NVIDIA's open-source framework for optimizing LLM inference with advanced quantization techniques.
We utilize the latest stable release of TensorRT-LLM (v0.20.0) for comparison.
\vspace{0.25em}
\item \textbf{OmniServe+QServe.} OmniServe+QServe~\cite{lin2024qserve} is a serving system specifically optimized for \seqsplit{W4A8KV4} quantization format. 
We utilize the latest release for comparison.
\end{itemize}

\vspace{0.5em}
\noindent\textbf{Models and workloads.} 
Our experiments include models from the Qwen, Llama, DeepSeek, and Mixtral series, spanning different sizes (8B, 32B, 70B, and 235B parameters) and different quantization methods (AWQ~\cite{lin2024awq} and GPTQ~\cite{frantar2022gptq}). For general conversational tasks, we evaluate these models using real-world chatbot workloads derived from the ShareGPT dataset~\cite{sharegpt2023}. For reasoning tasks, we conduct experiments using the reasoning model QwQ~\cite{qwq32b} on mathematical reasoning workloads from the NuminaMath dataset~\cite{numinamath_cot} and the AIMO validation dataset~\cite{aimo_validation_aime}.  We follow prior works~\cite{li2023alpaserve,jiang2024hexgen} to generate the inference workload using a Poisson process determined by the request rate.

\vspace{0.5em}
\noindent\textbf{Evaluation metrics.} Following prior works~\cite{patel2024splitwise,frantar2025marlin,zhong2024distserve}, we focus on three primary performance metrics. First, we measure overall system throughput under different inference batch sizes and online serving scenarios. Second, we evaluate system response latency across various percentiles (P50, P90, P95, $\cdots$, P99), where P90 latency represents the maximum response time within which 90\% of all requests are completed. Third, we report the time-to-first-token (TTFT) latency, the elapsed time between receiving a request and emitting the very first output token, which directly affects perceived responsiveness in interactive applications.

\subsection{End-to-end Performance}
\label{sec:e2e}
\begin{figure}[t!]
    \centering
    \includegraphics[width=\linewidth]{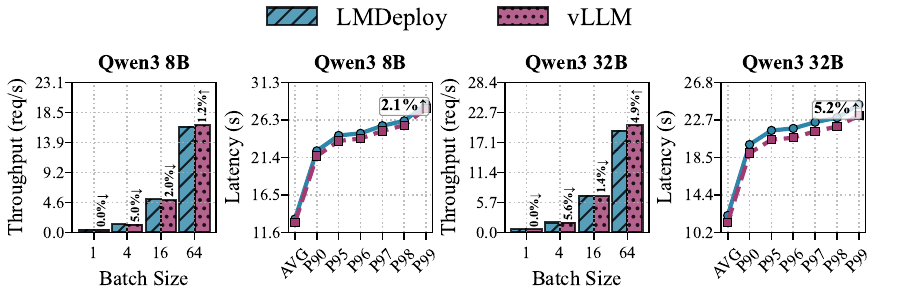}
    \caption{Latency comparison between \sys and vLLM using general inference configuration W16A16KV16 (without mixed-precision formats) on H100 GPUs.}
    \label{fig:general}
\end{figure}

\noindent\textbf{Compare \sys with vLLM+MARLIN.} As shown in~\autoref{fig:overall}, results show that \sys consistently outperforms vLLM+MARLIN~\cite{kwon2023efficient,frantar2025marlin} across all performance metrics and experimental configurations. We evaluated both systems using Qwen 8B and 32B AWQ models across four different GPU types (RTX 4090, L40, A100, and H100) under varying workload conditions. In terms of throughput, \sys achieves an average speedup of 13\% compared to vLLM+MARLIN, with a maximum speedup of 31\% observed under high-batch scenarios. For TTFT, \sys reduces latency by an average of 12.0\% (maximum: 33.3\%). In online serving scenarios, \sys demonstrates superior latency characteristics with an average improvement of 15.0\% (maximum: 24.6\%) across all latency percentiles (P90-P99). Under varying request arrival rates (1.0 to 10.0 req/s), \sys shows even more pronounced benefits, achieving an average latency reduction of 24.1\% with maximum improvements of 37.3\% at higher request rates. These consistent improvements demonstrate the robustness and general applicability of \sys's mixed-precision inference pipeline (specifically, GEMM pipeline in \S\ref{sec:pipelines}). We also compare \sys with vLLM under a general inference configuration. As shown in~\autoref{fig:general}, \sys performs 2.1\% and 5.2\% worse than vLLM under the general format (W16A16K16). This confirms that our performance gains originate from mixed-precision optimization rather than framework differences.

\begin{figure}[t!]
    \centering
    \includegraphics[width=\linewidth]{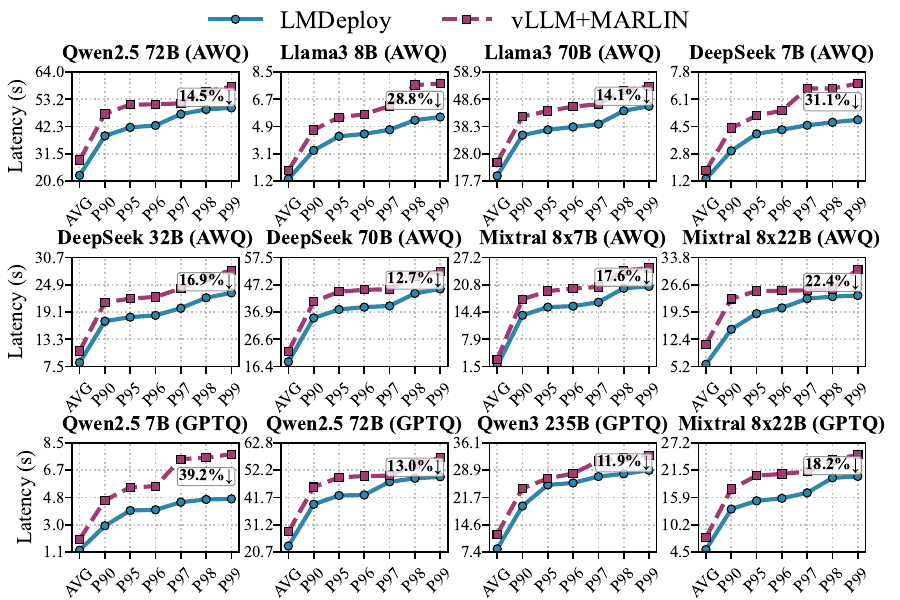}
    \caption{Serving latencies of \sys compared with vLLM+MARLIN on different models on A100 GPUs.}
    \label{fig:diff_model} 
\end{figure}

\vspace{0.5em}
\noindent \textbf{Compare \sys with vLLM+MARLIN across a wider range of models.} As shown in~\autoref{fig:diff_model}, 
we conducted an extensive evaluation across 12 diverse models spanning different architectures, parameter scales, and quantization methods. This broader evaluation includes dense models (Llama, Qwen, DeepSeek series) ranging from 7B to 235B parameters, as well as MoE models (Mixtral series), using both AWQ and GPTQ quantization techniques. Across this comprehensive model suite, \sys achieves an average serving latency improvement of 21.1\%, with maximum improvements reaching 47.9\%. 
At the critical P99 latency percentile, \sys delivers an average improvement of 20.0\% with peak improvements of 39.2\%, ensuring reliable performance even under tail latency conditions. These results establish \sys as a robust solution for production LLM serving across diverse contemporary LLMs.

\begin{figure}
    \centering
    \includegraphics[width=\linewidth]{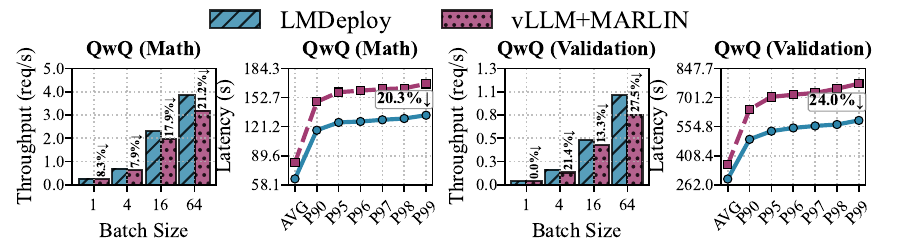}
    \caption{Latency and throughput comparison between \sys and vLLM+MARLIN on QwQ with math and validation workloads on an A100 GPU.}
    \label{fig:reasoning}
\end{figure}

\vspace{0.5em}
\noindent\textbf{Compare \sys with vLLM+MARLIN with reasoning models and workloads.} As shown in~\autoref{fig:reasoning}, to evaluate \sys's effectiveness on reasoning-intensive workloads, we conducted specialized evaluations using QwQ AWQ models designed for mathematical reasoning and validation tasks. For throughput performance, \sys achieves an average speedup of 15\% compared to vLLM+MARLIN, with peak improvements of 27\% observed in validation tasks. And \sys delivers an average latency reduction of 21.9\% across all percentiles, with maximum improvements reaching 24.5\%. At the critical P99 latency percentile, \sys maintains robust performance with 20.3\% improvement for mathematical reasoning tasks and 24.0\% improvement for validation workloads. These results establish \sys's suitability for sophisticated reasoning tasks.

\begin{figure}
    \centering
    \includegraphics[width=\linewidth]{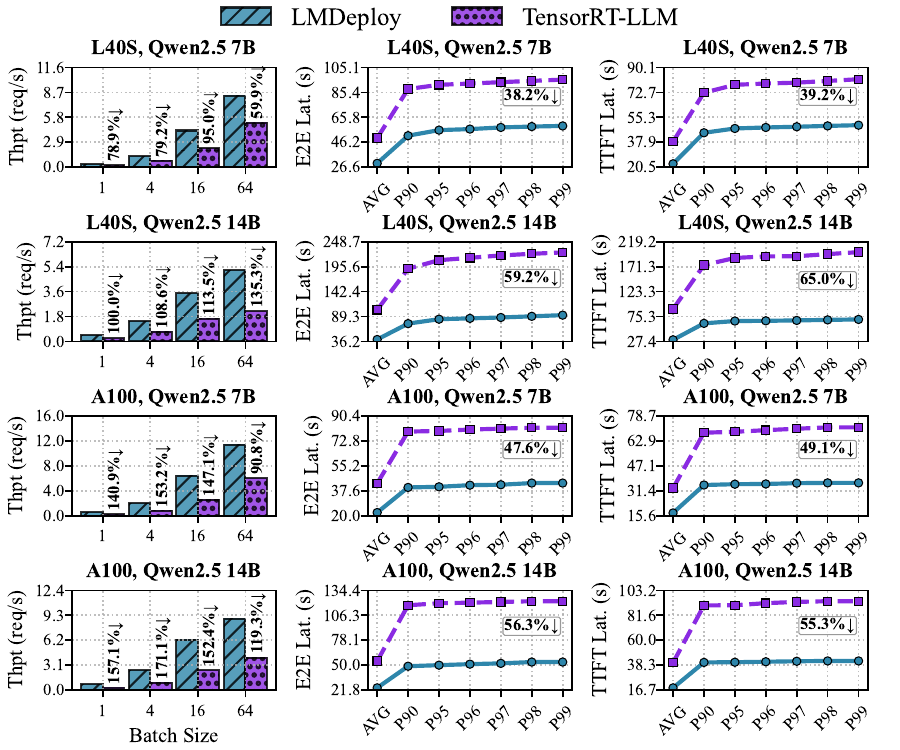}
    \caption{End-to-end experiments of \sys compared with TensorRT-LLM on L40S and A100 GPUs.}
    \label{fig:tensorrt}
\end{figure}

\vspace{0.5em}
\noindent\textbf{Compare \sys with TensorRT-LLM.} As shown in~\autoref{fig:tensorrt}, to demonstrate \sys's performance against NVIDIA's highly optimized inference engine TensorRT-LLM \cite{tensorrt-llm}, we conducted comparisons using Qwen 7B and 14B AWQ models across multiple evaluation metrics. In throughput performance, \sys achieves an average speedup of 118.90\%, with peak speedups reaching 171.11\% across different batch configurations. And \sys reduces TTFT by an average of 52.2\%, with maximum improvements of 65.0\%. For end-to-end latency across all percentile measurements, \sys delivers an average reduction of 50.3\%, with peak improvements of 59.2\%. These substantial performance gains further underscore the effectiveness of \sys.

\begin{figure}
    \centering
    \includegraphics[width=\linewidth]{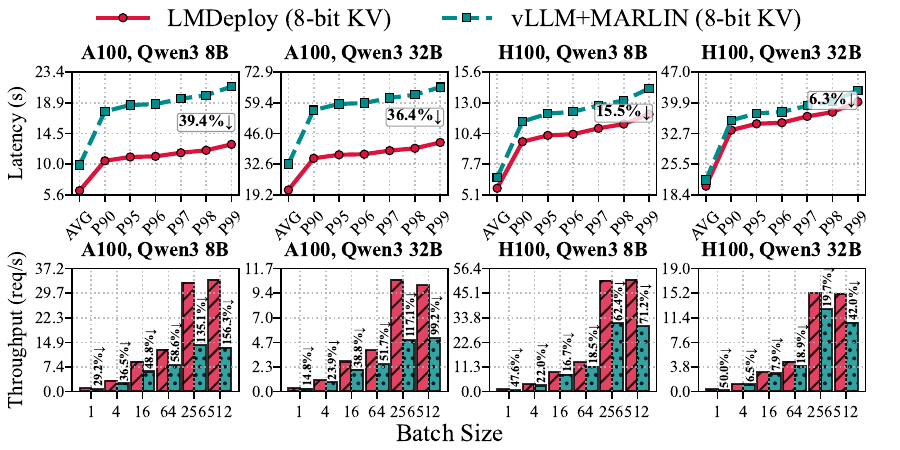}
    \caption{Latency and throughput comparison between \sys and vLLM+MARLIN with 8-bit KV cache.}
    \label{fig:kv}
\end{figure}

\vspace{0.5em}
\noindent\textbf{Compare \sys with vLLM+MARLIN with 8-bit KV cache.} As shown in~\autoref{fig:kv}, we conducted comprehensive evaluations comparing \sys with INT8 KV cache support against vLLM+MARLIN with FP8 KV cache support~\cite{vllm_quantized_kvcache}, utilizing Qwen 8B and 32B AWQ models on both A100 and H100 GPUs. For throughput performance, \sys achieves an average speedup of 50.6\% compared to vLLM+MARLIN, with peak speedups reaching 156.3\% in high-batch scenarios on A100 GPUs. 
And \sys reduces latency by an average of 24.6\% (maximum: 40.5\%) across all percentiles, with P99 latency improving by 24.4\% (maximum: 39.4\%).
These results demonstrate the effectiveness of \sys's attention pipeline (\S\ref{sec:pipelines}) in low-bit KV cache inference and serving. 

\noindent \textbf{Accuracy evaluation.} To verify the correctness of our low-bit KV cache implementation, we demonstrate the accuracy performance comparison between \sys and vLLM across three popular models and several existing datasets, as shown in~\autoref{tab:accuracy}. The evaluation results demonstrate the accuracy equivalence between the two systems: Both systems achieve highly comparable performance across diverse benchmark tasks, with average score differences consistently remaining within 1-4 percentage points, across multiple evaluation domains.

\begin{table}[h]                       
\centering
\caption{Performance comparison between \sys and vLLM using an 8-bit KV cache across various models and datasets.}
\setlength{\tabcolsep}{6pt}            
\renewcommand{\arraystretch}{1.1}      
\resizebox{\linewidth}{!}{%
\begin{tabular}{l|l|c|c|c|c}         
\hline
\textbf{Model} & \textbf{System} &
\textbf{GSM8K} & \textbf{MMLU} & \textbf{MMLU-STEM} & \textbf{MMLU-Hum} \\
\hline
\multirow{2}{*}{Qwen 14B} & vLLM & 84.23 & \textbf{87.30} & \textbf{92.36} & 83.42 \\
\cline{2-6}
                           & \sys & \textbf{84.38} & 87.29 & 92.05 & \textbf{83.94} \\
\hline
\multirow{2}{*}{Qwen 8B}  & vLLM & 76.12 & 85.24 & 89.18 & 81.31 \\
\cline{2-6}
                           & \sys & \textbf{80.97} & \textbf{85.59} & \textbf{89.93} & \textbf{81.60} \\
\hline
\multirow{2}{*}{Llama 8B} & vLLM & 80.67 & 65.82 & 62.84 & 65.12 \\
\cline{2-6}
                           & \sys & \textbf{81.20} & \textbf{71.40} & \textbf{67.05} & \textbf{72.61} \\
\hline
\end{tabular}}
\label{tab:accuracy}
\end{table}

\begin{figure}[t!]
    \centering
    \includegraphics[width=\linewidth]{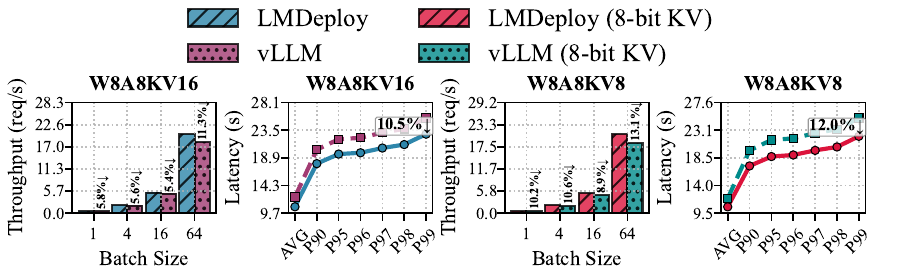}
    \caption{Latency and throughput comparison between \sys and vLLM+MARLIN with the FP8 Qwen3 8B model on an H100 GPU.}
    \label{fig:fp8_test}
\end{figure}

\vspace{0.5em}
\noindent\textbf{Compare \sys with vLLM on FP8 models.} As shown in~\autoref{fig:fp8_test}, to further evaluate \sys's effectiveness on FP8 quantized models with different KV cache precision levels, we conducted additional evaluations. Compared with the baseline approach, \sys achieves an average performance improvement of 10.4\% across both throughput and latency metrics, with peak improvements reaching 13.1\%. The consistent performance improvements 
further demonstrate \sys's comprehensive support for different mixed-precision inference cases.

\begin{figure}[t!]
    \centering
    \includegraphics[width=\linewidth]{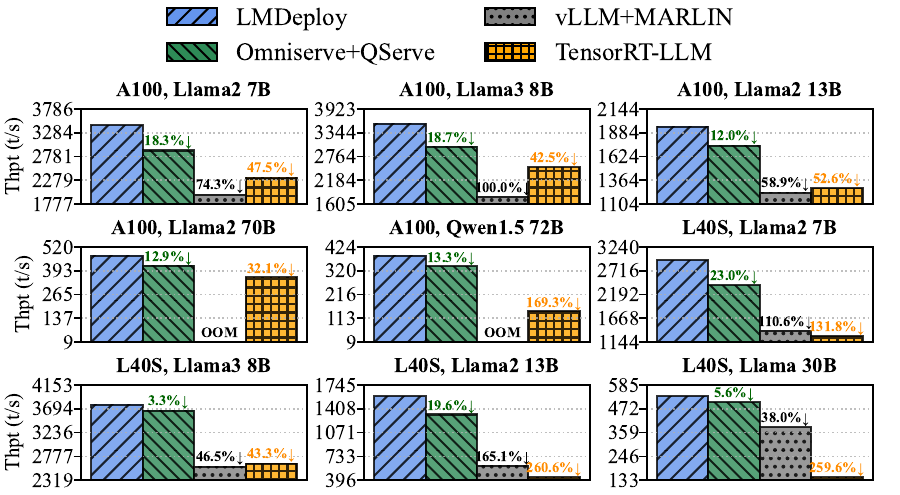}
    \caption{Maximum throughput comparison among \sys and baseline frameworks. The experiments follow the benchmarking setting in QServe~\cite{lin2024qserve}.}
    \label{fig:qserve}
\end{figure}

\vspace{0.5em}
\noindent\textbf{Compare \sys with OmniServe+QServe.} As shown in~\autoref{fig:qserve}, we compare four systems with their optimal precision formats: (\textbf{\underline{i}}) our system \sys (W4A16KV4); (\textbf{\underline{ii}}) OmniServe+QServe (W4A8KV4)~\cite{lin2024qserve}, which is a system specifically optimized for W4A8KV4 mixed-precision LLM serving; (\textbf{\underline{iii}}) vLLM+MARLIN (W4A8KV8); and (\textbf{\underline{iv}}) TensorRT-LLM (W16A16/W4A16/W8A8KV16), wherein we systematically evaluate the available precision formats across different GPU and model configurations and report the optimal variant for each case.
\sys consistently outperforms all baselines, achieving an average throughput improvement of 14.1\% (maximum: 23.0\%) over OmniServe-QServe despite the latter's use of more aggressive 8-bit activation quantization, 84.8\% (maximum: 165.1\%) over \seqsplit{vLLM+MARLIN}, and 115.5\% (maximum: 260.6\%) over TensorRT-LLM across different GPUs and models. Results show \sys's superior performance even compared to frameworks specifically engineered for certain mixed-precision formats.

\begin{figure}[t!]
    \centering
    \includegraphics[width=\linewidth]{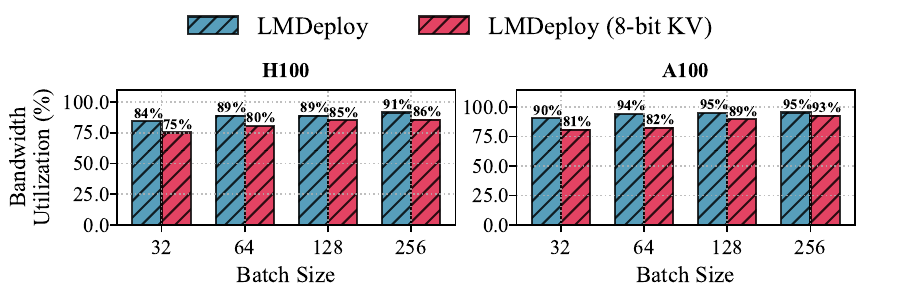}
    \caption{Memory bandwidth utilization of \sys’s attention kernel at different batch sizes.}
    \label{fig:bwutil}
\end{figure}

\subsection{Kernel Performance and Microbenchmark}
\label{sec:kernelbench}

\begin{figure}[t!]
    \centering
    \includegraphics[width=\linewidth]{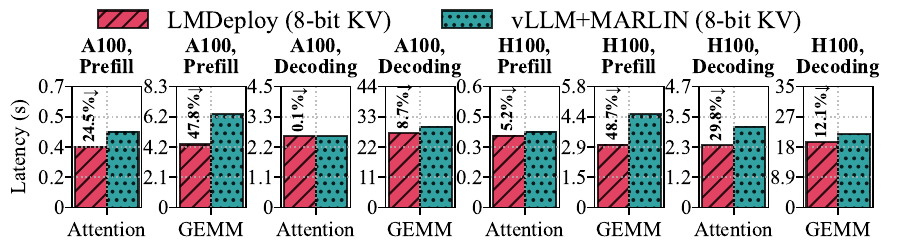}
    \caption{Benchmarking results of prefill and decoding latency for attention and GEMM kernels within a single request on the Qwen3 8B AWQ model with 8-bit KV cache.}
    \label{fig:prefilldecode}
\end{figure}

\begin{figure}[t!]
    \centering
    \includegraphics[width=\linewidth]{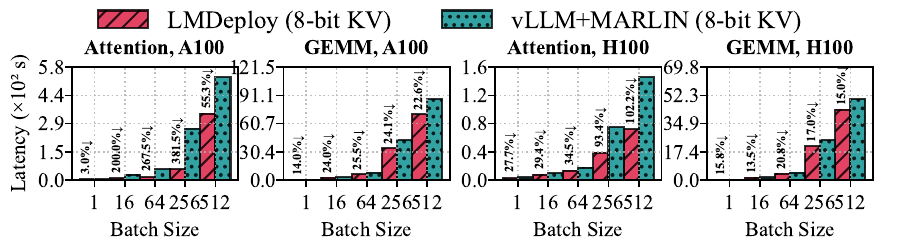}
    \caption{Benchmarking results of accumulated attention and GEMM kernel execution latencies on the Qwen3 8B AWQ model with 8-bit KV cache.}
    \label{fig:attngemm}
\end{figure}

\noindent\textbf{Compare our attention and GEMM kernel performance with vLLM+MARLIN.} As shown in~\autoref{fig:prefilldecode} and~\autoref{fig:attngemm}, to evaluate the effectiveness of our GEMM and attention pipeline/kernel design in~\S\ref{sec:pipelines}, we implement a \seqsplit{micro-benchmark} evaluation on the GEMM and attention kernels using the Qwen 8B AWQ model with 8-bit KV cache (a mixed-precision format of W4A16KV8). We choose vLLM+MARLIN with 8-bit KV cache compression (\texttt{fp8\_e5m2}~\cite{vllm_quantized_kvcache}) as the baseline method. The results demonstrate that our optimized attention kernel achieves average latency reductions of 22.1\% (maximum: 48.7\%) during prefill operations and 7.6\% (maximum: 29.9\%) during decode operations compared with the baseline method, and deliver average throughput improvements of 88.5\% (maximum: 381.7\%) across varying batch sizes, while our optimized GEMM kernels demonstrate average performance gains of 19.2\% (maximum: 25.5\%) compared to the baseline method. These empirical results substantiate the significance of our GEMM and attention pipeline/kernel design, establishing \sys's superior per-kernel performance relative to the state-of-the-art solution. We also demonstrate the memory bandwidth utilization of \sys's attention kernel. As shown in~\autoref{fig:bwutil}, \sys's attention kernel achieves outstanding memory bandwidth utilization across different batch sizes, reaching up to 91\% and 95\% with 16-bit KV cache, and achieving a maximum of 86\% and 93\% with 8-bit KV cache, further proving the effectiveness of our attention pipeline design (\S\ref{sec:pipelines}).

\begin{figure}[t!]
    \centering
    \includegraphics[width=\linewidth]{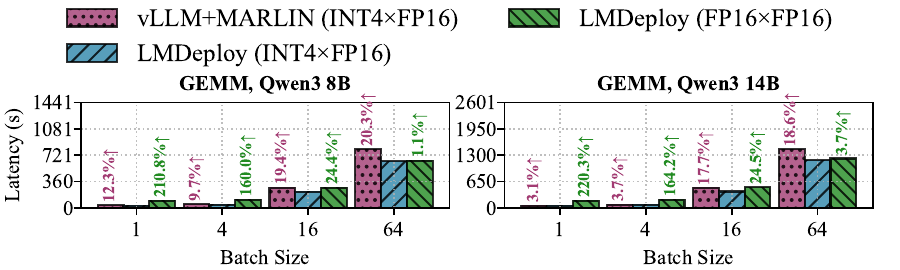}
    \caption{Benchmarking results of our INT4$\times$FP16 kernel versus a general FP16$\times$FP16 GEMM kernel on an A100 GPU.}
    \label{fig:int4}
\end{figure}

\vspace{0.5em}
\noindent\textbf{Compare our mixed-precision GEMM kernel with general GEMM kernel.} As illustrated in~\autoref{fig:int4}, to evaluate the effectiveness of our instruction-level parallelism mentioned in~\S\ref{sec:ilp}, we further compare our \seqsplit{INT4$\times$FP16} GEMM kernel against the general \seqsplit{FP16$\times$FP16} GEMM kernel.
For small-batch configurations (batch sizes 1-16), our INT4$\times$FP16 kernel achieves an average latency improvement of 134\% (maximum 220.3\%) over FP16$\times$FP16 implementations. For large-batch scenarios (batch size 64), our INT4$\times$FP16 kernel maintains performance parity with general FP16$\times$FP16 kernels, while the MARLIN mixed-precision kernel suffers up to 20.3\% performance degradation. These results validate our instruction-level parallelism design for achieving satisfactory performance during dequantization operations.

\begin{table}[htbp]
\centering
\caption{Performance comparison between \sys (INT4$\times$FP16) and cuBLAS (FP16$\times$FP16) GEMM kernels at full utilization on an A100 GPU (problem size: $16384^3$).}
\resizebox{\linewidth}{!}{%
\begin{tabular}{@{}l | c | c@{}}
\hline
 & \textbf{\sys (INT4$\times$FP16)} & \textbf{cuBLAS (FP16$\times$FP16)} \\
\hline
Instr. Count & $7,145,914,386$ $(64.66\%\uparrow)$ & $4,339,924,992$ \\
\hline
Cycle Count & $41,864,631$ $(2.89\%\uparrow)$ & $40,690,070$ \\
\hline
Runtime (ms) & $30.28$ $(2.45\%\uparrow)$ & $29.55$ \\
\hline
\end{tabular}
}
\label{tab:performance}
\end{table}

We also compare our \seqsplit{INT4$\times$FP16} against cuBLAS's \seqsplit{FP16$\times$FP16} GEMM kernel in terms of instruction and cycle counts. The results (shown in~\autoref{tab:performance}) demonstrate that our kernel requires 64.66\% more instructions than cuBLAS's general kernel, primarily due to the additional dequantization operations. However, this instruction overhead translates to only 2.89\% more cycles and 2.45\% longer execution time, further demonstrating the effectiveness of our instruction-level parallelism in hiding the dequantization latency.

\vspace{0.5em}
\noindent\textbf{Microbenchmark on other techniques.} Adaptive head alignment (\S\ref{sec:adaptive-head}) occurs twice per decoding step (Q initialization and output write-back), incurring only microsecond-level overhead per attention head ($<$1\% of total attention latency). This overhead is negligible compared to naïve K dequantization to global memory (3$\times$ memory footprint, $>$50\% performance degradation). Hardware-aware weight packing (\S\ref{sec:hardware-aware}) completes in seconds even for large models (e.g., 1.47s for 70B parameters) and is performed only once before serving, introducing negligible preprocessing cost.

\begin{figure}
    \centering
    \includegraphics[width=\linewidth]{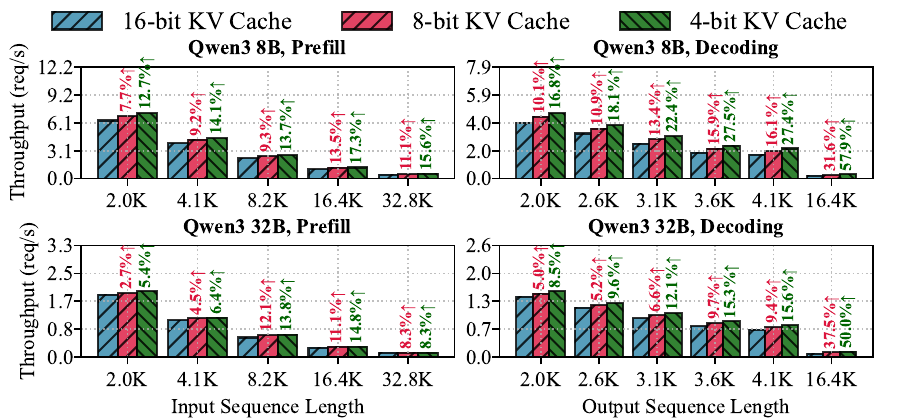}
    \caption{Throughput comparison between different KV precision of \sys with different serving batch sizes on an A100 GPU.}
    \label{fig:kv_precision}
\end{figure}

\subsection{Sensitivity and Scalability}

\noindent\textbf{\sys performance with different KV cache precisions.} As shown in~\autoref{fig:kv_precision}, \sys demonstrates exceptional performance scalability across different KV cache precision levels. Our comprehensive evaluation across Qwen 8B and 32B AWQ models reveals consistent throughput improvements as KV cache precision is reduced, with benefits observed across both prefill and decoding phases at diverse sequence lengths. When utilizing 8-bit KV cache quantization compared to the 16-bit baseline, \sys achieves an average throughput improvement of 11.9\%, with peak gains reaching 37.5\% in long sequence scenarios. The performance advantages become even more pronounced with aggressive 4-bit KV cache quantization, where \sys delivers an average improvement of 18.3\% over the 16-bit baseline, with maximum speedups of 57.9\% observed in long sequence scenarios. These results demonstrate the efficiency benefits of KV cache compression, while further validating the significance of \sys's attention pipeline design (\S\ref{sec:pipelines}). 

\begin{figure}[t!]
    \centering
    \includegraphics[width=\linewidth]{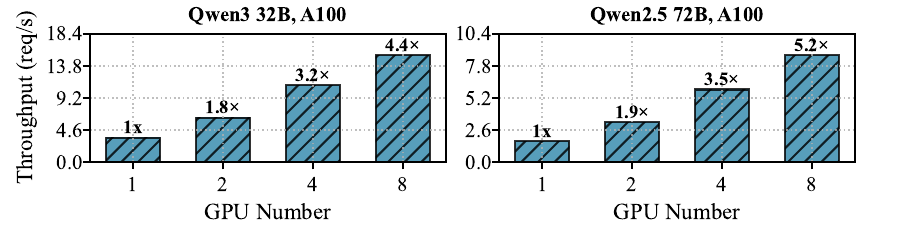}
    \caption{Scalability of \sys in multi-GPU serving (tensor parallelism degree = \{1, 2, 4, 8\}).}
    \label{fig:scale}
\end{figure}

As shown in~\autoref{fig:scale}, we also demonstrate the scalability of \sys with increased tensor parallelism degree. For Qwen 32B and Qwen 72B AWQ models, \sys scales from 3.44 req/s and 1.68 req/s on a single GPU to 15.3 req/s and 8.7 req/s on 8 GPUs, representing a 4.45$\times$ and 5.18$\times$ improvement with 55.6\% and 64.8\% parallel efficiency, respectively. These results highlight \sys's ability to effectively leverage distributed computing resources for high-throughput LLM serving.

\section{Future Work}
\label{sec:futurework}

While \sys with \eng focuses on generalizable and efficient mixed-precision inference through hardware-aware GEMM and attention pipelines, several directions remain for further extending its applicability and deployment efficiency.

\vspace{0.5em}
\noindent\textbf{Accuracy-aware precision selection.}
The current design uses fixed mixed-precision configurations during inference, but it can be extended with runtime policies that adapt KV cache precision across models, layers, and workload conditions. Such policies could assign lower precision to accuracy-tolerant components while preserving higher precision for sensitive layers, heads, or requests. This would allow \eng to further improve serving efficiency while maintaining model quality.

\vspace{0.5em}
\noindent\textbf{Distributed mixed-precision serving.}
The current system-level optimizations can also be extended to large-scale distributed serving environments. In such deployments, precision selection can be jointly optimized with data/model parallelism, request scheduling, memory allocation, and cost-efficiency objectives. Exploring such system-level co-\seqsplit{optimization} could further extend the benefits of \eng beyond single-node inference and enable more efficient mixed-precision LLM serving at cluster scale.

\section{Conclusion}
\label{sec:conclusion}
This paper presents the mixed-precision LLM inference and serving techniques integrated in \sys that systematically address memory and compute optimization challenges through novel GEMM and attention pipelines implemented in the \eng engine.
Our evaluation demonstrates consistent performance improvements,
achieving 12-61\% lower serving latency with 13-156\% higher throughput compared to state-of-the-art mixed-precision frameworks. These results establish \sys as an efficient solution for mixed-precision LLM deployment, enabling broader accessibility of LLMs through optimized resource utilization.

\bibliographystyle{ACM-Reference-Format}
\bibliography{main}

\newpage
\appendix
\onecolumn

\section{Illustration of Challenges}
\label{appendix:challenges}

\begin{figure}[htbp]
    \centering
    \includegraphics[width=0.6\linewidth]{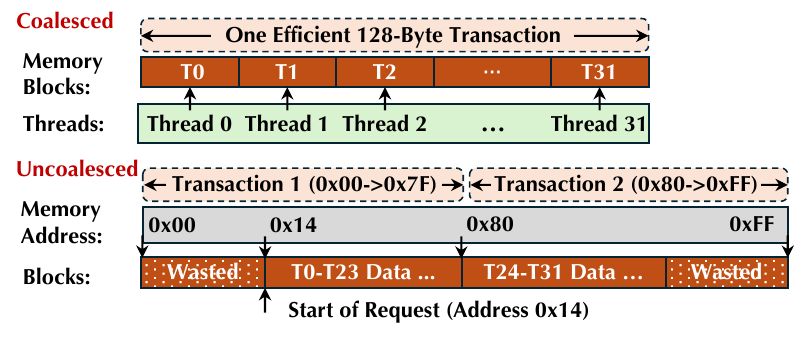}
    \caption{Illustration of wasted memory bandwidth due to uncoalesced memory access (an example of two transactions).}
    \label{fig:globalmem}
\end{figure}

\noindent Modern GPUs achieve peak memory bandwidth when the memory addresses accessed by every thread within a warp are within the same aligned segment of global memory (e.g., 3-byte on Hopper/Ampere). This alignment enables the warp to access contiguous memory regions through one efficient global memory transaction~\cite{alur2017gpudrano,fauzia2015characterizing,kim2017evaluation}.
In mixed-precision inference, however, packing weights into low-bit formats results in misalignment between each warp's memory accesses and the GPU's standard 32-/64-/128-byte memory segments~\cite{cuda_prog_guide_128b_2024,pytorch_gptq_optimization}. This misalignment necessitates multiple global memory transactions per warp for operations rather than one efficient transaction, 
thereby significantly reducing effective memory bandwidth~\cite{kim2024quick,frantar2022gptq}. As shown in~\autoref{fig:globalmem}.

\begin{figure}[t!]
    \centering
    \includegraphics[width=0.6\linewidth]{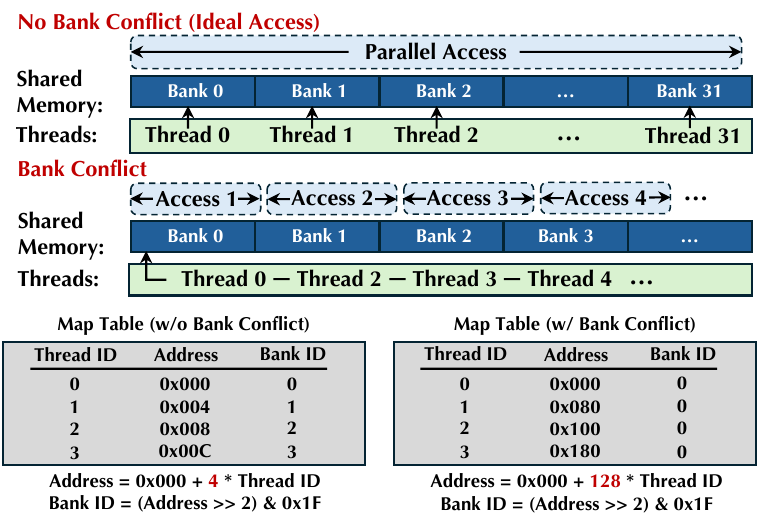}
    \caption{Illustration of reduced memory throughput due to shared memory bank conflicts (an example of 32-way bank conflict).}
    \label{fig:sharedmemory}
\end{figure}

\noindent Shared memory bank conflicts occur when multiple threads within a warp simultaneously access different addresses that map to the same memory bank, forcing these accesses to be serialized rather than executed in parallel, thereby reducing memory throughput by up to an order of magnitude~\cite{biswas2020memory,horga2022symbolic,gao2014improving,lin2024qserve}. In mixed-precision inference, low-bit values are typically packed into larger-bit words (e.g., eight INT4 values per 32‑bit word~\cite{nvidia_tensorrt_quant}) to achieve optimal memory bandwidth and space efficiency~\cite{frantar2025marlin,rakka2022mixed,pytorch,tensorrt-llm}. 
However, packing low‑bit weights into larger-bit words makes each column load jump a full packed row (e.g., 128-byte for 32 threads read 32‑bit words)~\cite{cuda_prog_guide_128b_2024}. That stride maps every thread in the warp to the same shared memory bank, so the bank must serve the requests one‑by‑one, turning a parallel-access load into a serialized‑access stall, thereby severely degrading memory throughput. As shown in~\autoref{fig:sharedmemory}.

\begin{figure}[t!]
    \centering
    \includegraphics[width=0.6\linewidth]{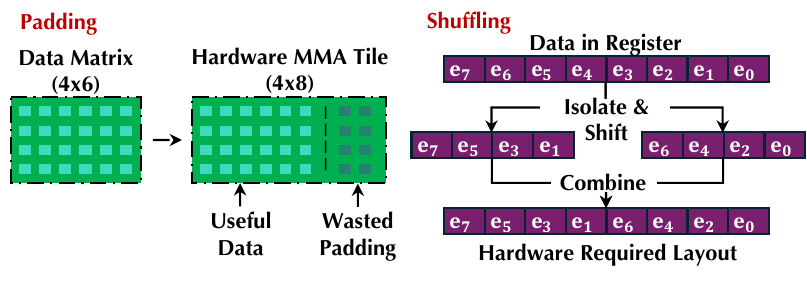}
    \caption{Illustration of padding and shuffling in MMA data misalignment.}
    \label{fig:mma}
\end{figure}

\noindent Modern tensor cores provide INT8/INT4 MMA instructions, but they only run at full throughput when the inputs are pre-packed into fixed tiles (e.g., 16$\times$8$\times$32/64 for Ampere, 16$\times$8$\times$64/128 for Hopper)~\cite{nvidia_ampere_tuning_guide,luo2024benchmarking}.
Standard quantization layouts seldom meet these hardware requirements, creating a fundamental mismatch between them. Consequently, kernels must either inject padding that wastes compute resources, perform costly in-register shuffles at runtime (\autoref{fig:mma}), 
or fall back to less efficient scalar operations—all of which erode tensor core throughput \cite{lin2024awq,frantar2025marlin}.

\section{Swizzling}
\label{appendix:swizzle}
\begin{figure}[htbp]
    \centering
    \includegraphics[width=\linewidth]{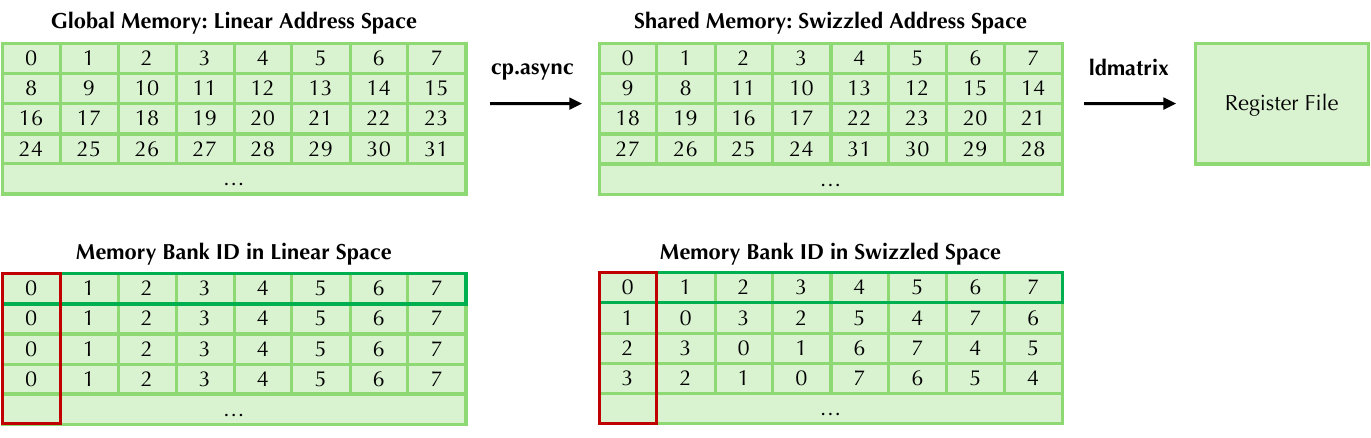}
    \caption{Illustration of 8$\times$128 byte swizzle unit.}
    \label{fig:swizzle}
\end{figure}

\textbf{Why swizzling operations are required?} \autoref{fig:swizzle} illustrates why a swizzled shared memory layout is used in mixed‑precision inference. During the prefetch phase, \texttt{cp.async} issues horizontal, coalesced row writes to shared memory. During the compute phase, \texttt{ldmatrix} performs vertical, per‑lane column reads (in groups of eight threads) to feed the tensor cores. With a naïve row‑major layout, those vertical reads cause multiple lanes to hit the same shared memory bank, leading to bank conflicts or extra shuffles. After swizzling, the same logical tile is permuted so that the \texttt{ldmatrix} loads are conflict‑free and each lane receives exactly the elements the MMA instruction expects, while the horizontal \texttt{cp.async} writes remain coalesced.

\noindent \textbf{Why does our packing approach avoid swizzling?} Our packing approach packs each logical 16$\times$16 tile into a 32$\times$8 layout; this essentially bakes the swizzle in offline. The 32 rows are just four 8$\times$8 sub‑tiles stacked vertically, so when \texttt{ldmatrix}'s 8 participating lanes read columns (vertical), each lane walks its own column and lands on distinct shared memory banks, delivering exactly the fragments tensor cores expect without any runtime shuffles or bank conflicts. Meanwhile, \texttt{cp.async} can still issue horizontal, coalesced row writes because every 32‑element row is contiguous in memory.

\section{Rearrangement Algorithm}
\label{appendix:rearrangement}

Algorithm~\autoref{alg:transformQ} demonstrates the rearrangement procedure for the \texttt{m16n8k16} tensor core instruction with a KV head dimension of 128. The variable $X$ is defined as 16 divided by the bit-width of the KV parameter. For example, $X$ equals 2 for an 8-bit KV and 4 for a 4-bit KV.

\begin{algorithm}[t]
\caption{Q Rearrangement Algorithm}
\label{alg:transformQ}
\begin{algorithmic}[1]
\REQUIRE{$Q_{\text{sm}}$: Query tile in shared memory, $\text{HeadDim}$: attention head dimension, $\text{OP\_K}$: tensor core operand granularity, $\text{OP\_N}$: N-dimension operand size, $K_N$: warp tile size along N dimension, $X$: batch processing size}
\ENSURE{Register fragment $\mathsf{frag\_Q}$ in \texttt{m16n8k16} layout}
\STATE \textcolor{darkcyan}{/* Thread coordinates within block */}
\STATE $lane \leftarrow \text{lane\_id}$ 
\STATE \textcolor{darkcyan}{/* Step 1: Compute K-slices based on precision */}
\STATE $K_K \leftarrow \text{HeadDim} / \text{OP\_K}$
\STATE \textcolor{darkcyan}{/* Iterate over N dimension, K-slices with batching */}
\FOR{$n = 0$ {\bfseries to} $K_N-1$}
    \FOR{$k = 0$ {\bfseries to} $K_K-1$ \textbf{step} $X$}
        \FOR{$x = 0$ {\bfseries to} $X-1$}
            \STATE \textcolor{darkcyan}{/* Load two fragments at once */}
            \FOR{$d = 0$ {\bfseries to} $1$}
                \STATE \textcolor{darkcyan}{/* Step 2: Coordinate thread mapping to Q */}
                \STATE $hi \leftarrow n \cdot \text{OP\_N} + \lfloor lane/4 \rfloor$ 
                \STATE $di \leftarrow k \cdot \text{OP\_K} + (lane \bmod 4) \cdot 2X + 2x + 8dX$ 
                \STATE \textcolor{darkcyan}{/* Step 3: Load and rearrange Q */}
                \STATE $\mathsf{frag\_Q}[n][k+x][2d] \leftarrow \texttt{Load}(Q_{\text{sm}}[hi][di])$ 
            \ENDFOR
        \ENDFOR
    \ENDFOR
\ENDFOR 
\STATE {\bfseries return} $\mathsf{frag\_Q}$ 
\end{algorithmic}
\end{algorithm}

\end{document}